# Equity Premium in Efficient Markets

B.N. Kausik[1]

Nov 19 2023

## Abstract

Equity premium, the surplus returns of stocks over bonds, has been an enduring puzzle. While numerous prior works approach the problem assuming the utility of money is invariant across contexts, our approach implies that in efficient markets the utility of money is polymorphic, with risk aversion dependent on the information available in each context, i.e. the discount on each future cash flow depends on all information available on that cash flow. Specifically, we prove that in efficient markets, informed investors maximize return on volatility by being risk-neutral with riskless bonds, and risk-averse with equities, thereby resolving the puzzle. We validate our results on historical data with surprising consistency.

---

[1] Conflict disclosure: unaffiliated independent.
Author's bio: https://sites.google.com/view/bnkausik/ Contact: bnkausik@gmail.com
Thanks to K. Suresh for comments suggestions

# Introduction

The surplus returns of stocks over bonds has been an enduring puzzle.

Mehra and Prescott (1985) posed the puzzle as a stochastic model in the Arrow & Debreu (1954) framework. In the model, the economy is a Markov chain where each state represents a fixed growth rate in per capita consumption. In its simplest form the model has two states, a growth state where per capita consumption enjoys above average growth, and a contraction state where per capita consumption suffers below average growth. Assuming a Constant Relative Risk Aversion utility function for the representative investor of the stand-in household, the model yields algebraic formulations for the returns of equities and riskless bonds. However, when model parameters are fit to the historical data on consumption growth, there are no reasonable values of the Coefficient of Constant Relative Risk Aversion or the Time Discount Factor that explains the surplus returns of stocks over bonds.

Since then, many works have attempted to resolve the puzzle, as reviewed by several authors such as Duarte & Rosa (2015), Mehra (2015) and Siegel (2017). The lattermost divides the literature into three broad categories: (1) *bias in historical data*, e.g., Brown, Goetzmann, and Ross (1995) argue that the US stock market is an outlier, while Dimson, Marsh, and Staunton (2008) find the US equity premium is comparable to that of other countries. McGrattan and Prescott (2001) suggest the US equity premium is an anomaly linked to the dividend tax rate, while Bernstein (1997), and Fama & French (2002) suggest the equity risk premium overstates future expected returns; (2) *model limitations and improvements*, e.g., Rietz (1988), Barro (2005), suggest that rare events, such as a crash state added to the stochastic model can explain the equity premium. However, the crash state required is catastrophically worse than the Great Depression; and (3) *behavioral finance considerations,* e.g. Benartzi and Thaler (1995), Barberis, Huang, and Santos (2001), Barberis and Huang (2007), and Constantinides (1990). However, the puzzle in its simplest two-state form as posed by Mehra and Prescott (1985) remains unsolved.

Against this backdrop, our approach introduces the assumption that the market is efficient and the stand-in household makes informed investment decisions, yielding the following: (1) With riskless bonds, the stand-in household is aware the bonds are indeed riskless and therefore maximizes return on volatility by being risk-neutral with a Coefficient of Constant Relative Risk Aversion equal to zero, and a Time Discount Factor reflecting the return of riskless bonds; (2) With equities, the stand-in household uses all available information to maximize return on volatility operating on the efficient frontier per Markowitz (1952). (3) We present an algorithm for estimating model parameters and optimal returns on historical data. (4) As a measure of robustness, the model's optimal return on equities is surprisingly consistent across the periods 1889-1979 and 1960-2022.

# The Economic Model

Following Mishra and Prescott (1985), our economic model is a Markov chain with $n$ states, where each State $i$ represents a consumption growth rate $\lambda_i$. For simplicity, we work with the minimal model of $n = 2$ states, i.e., a growth state and contraction state where $\lambda_1 > \lambda_2$. The representative stand-in household has a Constant Relative Risk Aversion utility function of the form:

$$U(c, \alpha) = (c^{1-\alpha} - 1)/(1 - \alpha), \quad \alpha \geq 0. \tag{1}$$

The role of the utility function in the model is to discount future cash flows. Specifically, the discounted utility of a future unit cash flow to the stand-in household at level $c$ is

$$U'(c + 1, \alpha)/U'(c, \alpha) = (1 + 1/c)^{-\alpha}$$

As $\alpha$ increases, the household is more risk-averse, increasing the discount on future cash flows. Prior works treat the utility function as invariant, i.e., $\alpha$ is a fixed constant across all contexts, the same for risky stocks and riskless bonds, in the growth state and in the contraction state. In an efficient market, informed investors utilize all available information to make decisions. Hence, our model views the utility function as polymorphic, i.e., $\alpha$ can vary with the information available in each context. Specifically, $\alpha$ can be distinct between stocks and bonds. In short, the discount on each future cash flow depends on all information available on that cash flow.

The Markov transition matrix $\Phi = \{\Phi_{i,j}\}$ specifies the probability of transitioning from State $i$ to State $j$. Each row of $\Phi$ sums to 1. The price of equities $w_i$ in each state satisfies the system of equations

$$w_i = \beta \sum_{j=1}^{n} \Phi_{i,j} \lambda_j^{1-\alpha^e} (w_j + 1) \tag{2}$$

where $\alpha^e$ is the Coefficient of Constant Relative Risk Aversion for equities and $0 \leq \beta \leq 1$ is the Time Discount Factor. The return on equities when transitioning from State $i$ to State $j$ is given by

$$r_{i,j}^e = \lambda_j(w_j + 1)/w_i - 1 \tag{3}$$

The expected return in State $i$ is

$$R_i^e = \sum_{j=1}^{n} \Phi_{i,j} r_{i,j}^e \tag{4}$$

And the total expected return for equities is

$$R^e = \sum_{i=1}^{n} \pi_i R_i^e \tag{5}$$

where $\pi = \Phi^T \pi$ is the stationary distribution of the Markov chain.

The standard deviation of the return on equities is

$$\sqrt{\sum_{i=1}^{n} \sum_{j=1}^{n} \pi_i \Phi_{i,j} (r_{i,j}^e - R^e)^2} \tag{6}$$

For riskless bonds, the price $p_i^f$ in State $i$ is

$$p_i^f = \beta \sum_{j=1}^{n} \Phi_{i,j} \lambda_j^{-\alpha^f} \tag{7}$$

where $\alpha^f$ is the Coefficient of Constant Relative Risk Aversion for bonds. The return in State $i$ is

$$R_i^f = 1/p_i^f - 1 \tag{8}$$

and the overall return on riskless bonds is

$$R^f = \sum_{i=1}^{n} \pi_i R_i^f \tag{9}$$

Let $\xi$ denote the mean, $\delta$ the standard deviation and $\sigma$ be the serial correlation in consumption growth across the two states.

**Assumption 1:** The serial correlation in consumption growth is non-zero, as is the case historically.

# Results

Our first result is a lemma that serves as a building block for our main results.

**Lemma 1:** If the serial correlation of consumption growth is non-zero, $\Phi$ is not singular.

**Proof:** Suppose the Markov matrix $\Phi$ is singular, then its two rows are identical, i.e.,
$\Phi_{1,1} = \Phi_{2,1}$ and $\Phi_{1,2} = \Phi_{2,2}$

Therefore,
$$\pi_1 = \pi_1 \Phi_{1,1} + \pi_2 \Phi_{2,1} \implies \pi_1 = \pi_1 \Phi_{1,1} + \pi_2 \Phi_{1,1} \implies \pi_1 = \Phi_{1,1}(\pi_1 + \pi_2) = \Phi_{1,1}$$

Likewise $\pi_2 = \Phi_{2,2}$.

Now, the serial correlation
$$\sigma = (\pi_1 \Phi_{1,1}(\lambda_1 - \xi)(\lambda_1 - \xi) + \pi_1 \Phi_{1,2}(\lambda_1 - \xi)(\lambda_2 - \xi) +$$
$$\pi_2 \Phi_{2,1}(\lambda_2 - \xi))(\lambda_1 - \xi) + \pi_2 \Phi_{2,2}(\lambda_2 - \xi))(\lambda_2 - \xi))/\delta^2$$

Multiplying both sides of the above by $\delta^2$ and noting that by definition, the mean $\xi = \pi_1 \lambda_1 + \pi_2 \lambda_2$,

$$\sigma \delta^2 = \pi_1^2 (\lambda_1 - \pi_1 \lambda_1 - \pi_2 \lambda_2)^2 + 2\pi_1 \pi_2 (\lambda_1 - \pi_1 \lambda_1 - \pi_2 \lambda_2)(\lambda_2 - \pi_1 \lambda_1 - \pi_2 \lambda_2) +$$
$$\pi_2^2 (\lambda_2 - \pi_1 \lambda_1 - \pi_2 \lambda_2)^2$$

Substituting $\lambda_1 = (\pi_1 + \pi_2)\lambda_1$ and $\lambda_2 = (\pi_1 + \pi_2)\lambda_2$ we get

$$= \pi_1^2 (\pi_2 \lambda_1 - \pi_2 \lambda_2)^2 + 2\pi_1 \pi_2 (\pi_2 \lambda_1 - \pi_2 \lambda_2)(\pi_1 \lambda_2 - \pi_1 \lambda_1) + \pi_2^2 (\pi_1 \lambda_2 - \pi_1 \lambda_1)^2$$

$$= 0 \text{ since } \lambda_1 > \lambda_2.$$

Therefore, if the serial correlation $\sigma$ is non-zero, $\Phi$ cannot be singular.

---

**Theorem 1:** In an efficient market, the stand-in household maximizes return on volatility by operating at a Coefficient of Constant Relative Risk Aversion $\alpha^f = 0$, and the Time Discount Factor $\beta = 1/(1 + R^f)$, where $R^f$ is the expected return on riskless bonds.

**Proof:** In an efficient market, the stand-in household is aware that riskless bonds have a constant payout and zero variance on their return, thereby maximizing return on volatility. Therefore $p_i^f$ is a constant $p^f$ across all states. Substituting in (7), we get,

$$p_i^f = \beta \sum_{j=1}^{n} \Phi_{i,j} \lambda_j^{-\alpha^f} = p^f \qquad (10)$$

One solution for (10) is $\alpha^f = 0$. To see this, set $\alpha^f = 0$ and since $\Phi$ is a Markov matrix, $\sum_{j=1}^{n} \Phi_{i,j} = 1$ to get

$$p_i^f = \beta \sum_{j=1}^{n} \Phi_{i,j} \lambda_j^{-\alpha^f} = p^f = \beta$$

Therefore, the expected return on riskless bonds is

$$R^f = \sum_{i=1}^{n} \pi_i(1/p_i^f - 1) = \sum_{i=1}^{n} \pi_i(1/p^f - 1) = \sum_{i=1}^{n} \pi_i(1/\beta - 1) = (1/\beta - 1)$$

which implies $\beta = 1/(1 + R^f)$.

Now suppose there exists alternate $\alpha \neq 0$ and time discount factor $\gamma$, that result in a constant price $q^f$ in each state, thereby achieving zero variance. Let $[\lambda^{-\alpha}]$ denote the column vector $[\lambda_1^{-\alpha}, \lambda_2^{-\alpha}]$ and $[q^f/\gamma]$ denote the column vector $[q^f/\gamma, q^f/\gamma]$. We can write

$$p_i^f = \gamma \sum_{j=1}^{n} \Phi_{i,j} \lambda_j^{-\alpha} = q^f \Rightarrow \sum_{j=1}^{n} \Phi_{i,j} \lambda_j^{-\alpha} = q^f/\gamma \Rightarrow \Phi[\lambda^{-\alpha}] = [q^f/\gamma]$$

Since $\Phi$ is a Markov matrix, $[q^f/\gamma] = \Phi[q^f/\gamma]$. Therefore,

$$\Phi[\lambda^{-\alpha}] = [q^f/\gamma] = \Phi[q^f/\gamma] \Rightarrow \Phi([\lambda^{-\alpha}] - [q^f/\gamma]) = 0$$

By Lemma 1, $\Phi$ is not singular. Therefore $([\lambda^{-\alpha}] - [q^f/\gamma])$ is the null vector $[0]$, which implies $\lambda_1^{-\alpha} = \lambda_2^{-\alpha} = q^f/\gamma$. Since $\lambda_1 > \lambda_2$, it follows that $\alpha = 0$ and $q^f = \gamma$, contradicting the assumption that $\alpha \neq 0$.

Therefore, at equilibrium, the stand-in household operates at $\alpha^f = 0$ and $\beta = 1/(1 + R^f)$.

---

Theorem 1 shows that in an efficient market, the stand-in household uses all available information to maximize return on volatility by being risk-neutral to riskless bonds. Likewise, in the context of equities, the stand-in household can use all available information to maximize return on volatility, per Markowitz (1952).

**Theorem 2:** In an efficient market for equities, the stand-in household maximizes return on volatility, i.e.,

$$dR^e/d\sigma = (R^e - R^f)/\sigma, \tag{11}$$

where $R^e$ is the expected return on equities, $\sigma$ is the standard deviation thereof, and $R^f$ is the return on riskless bonds.

**Proof:** If the stand-in household violates (11), there exists a portfolio combining riskless bonds and equities offering higher returns at the same volatility as the stand-in household. Therefore, the stand-in household will not maximize return on volatility. A contradiction, and hence the result.

# Experimental Validation

Table 1 summarizes the data for the period 1889-1978 as reported in Mishra and Prescott (1985).

| Table 1: Bonds, Stocks and Consumption Growth (1889-1978) | | | | |
|---|---|---|---|---|
| Riskless Bonds Real Return | S&P 500 Real Return (arithmetic mean) | Mean Real Consumption Growth | Standard Deviation Real Consumption Growth | Serial Correlation Real Consumption Growth |
| 0.008 | 0.0698 | 0.0183 | 0.0357 | -0.14 |

Algorithm 1 below searches the space of Markov models that fit Table 1 to maximize Return on Volatility. For simplicity, the algorithm assumes the Mean Real Consumption Growth is also the median, and hence the stationary probability is [0.5,0.5].

---

**Algorithm 1**

Let $\beta = 1/(1 + R^f)$ from Theorem 1
**for** probability distribution $\pi = [0.5, 0.5]$:
    compute $\lambda_1$, $\lambda_2$ to fit mean & standard deviation of consumption growth
    compute Markov transition matrix $\Phi$ to fit serial correlation
    **for** $\alpha^e \geq 0$:
        compute Return on Volatility $(R^e - R^f)/\sigma$ via Equations (5) and (6)
**Output** $\alpha^e$ that maximizes Return on Volatility

---

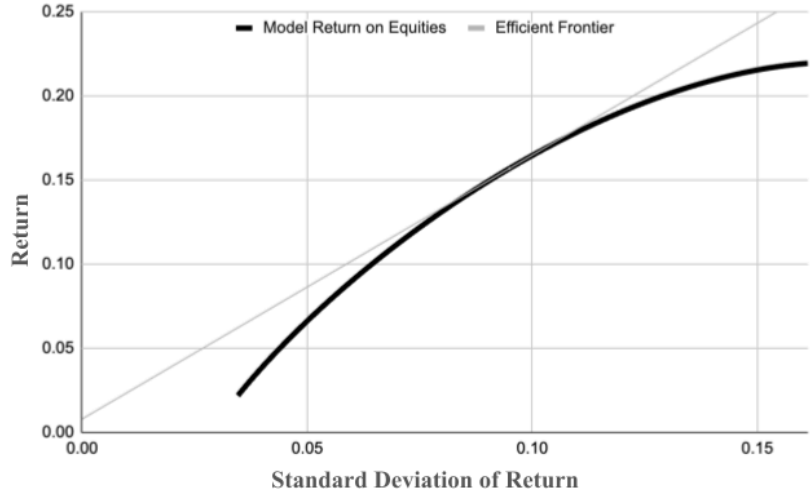

Fig 1. Return on Volatility (1889-1979)

Fig. 1 shows the Return on Volatility per Algorithm 1 on the data of Table 1 from 1889-1979. The bold curve is the Return on Equities vs Standard Deviation thereof for various values of $\alpha^e$. The point (0,0.008) represents the return of riskless bonds at zero standard deviation. The thin line in Fig. 1 is the tangent from the point (0,0.008) to the curve, thereby satisfying Equation (11) of Theorem 2. Specifically, at $\alpha^e \approx 9.75$, the tangent line forms the Efficient Frontier maximizing Return on Volatility for the stand-in household. The return on equities at the tangent point is 0.157, compared to the actual of 0.0698 in Table 1, suggesting the real world is less than efficient. The model achieves the actual return of 0.0698 at $\alpha^e \approx 3.5$. Any point $(v, R)$ on the tangent line represents a mix of riskless bonds and equities delivering the maximum return $R$ at volatility $v$, across all such portfolios. We now extend the timeline beyond the period 1889-1979 considered in Mehra and Prescott (1985). Specifically, from the data sources listed in Table 2a, we construct the statistics of Table 2b.

| Table 2a   Data Sources (1960-2022) | Source |
|---|---|
| Real S&P 500 Index and Dividends | Damodaran (2023) |
| Inflation, consumer prices for the United States | FRED (2023a) |
| Real personal consumption expenditures per capita: Services | FRED (2023b) |
| Real personal consumption expenditures per capita: Goods: Nondurable goods | FRED (2023c) |
| Market Yield on US Treasury Securities at 1-Year Constant Maturity | FRED (2023d) |

| Table 2b: Bonds, Stocks and Consumption Growth (1960-2022) | | | | |
|---|---|---|---|---|
| Riskless Bonds Real Return | S&P 500 Real Return (arithmetic mean) | Mean Real Consumption Growth | Standard Deviation Real Consumption Growth | Serial Correlation Real Consumption Growth |
| 0.0097 | 0.0733 | 0.0194 | 0.0158 | 0.15 |

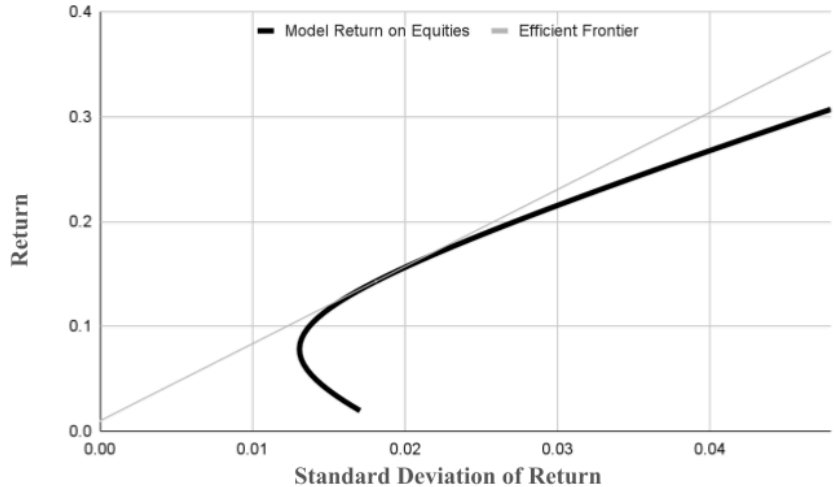

Fig 2. Return on Volatility (1960-2022)

Fig. 2 shows the Return on Volatility per Algorithm 1 on the data of Table 2b for 1960-2022. The bold curve is the Return on Equities vs Standard Deviation thereof for various values of $\alpha^e$. The point (0,0.0097) represents the return of riskless bonds at zero standard deviation. The thin line in Fig. 2 is the tangent from the point (0,0.0097) to the curve, thereby satisfying Equation (11) of Theorem 2. As in Fig. 1, at $\alpha^e \approx 6.75$, the tangent line forms the Efficient Frontier maximizing Return on Volatility for the stand-in household. The return on equities at the tangent point is 0.143, compared to the actual of 0.0733 in Table 2b, suggesting the real world is less than efficient. The model achieves the actual return of 0.0733 at $\alpha^e \approx 3.25$. Any point $(v, R)$ on the tangent line represents a mix of riskless bonds and equities delivering the maximum return $R$ at volatility $v$, across all such portfolios. The optimal return estimated by the model is surprisingly consistent across the periods 1889-1980 and 1960-2022, despite the substantial differences in statistics.

Our solution to the equity premium puzzle shows that in efficient markets, informed investors operate at different values of the Coefficient of Constant Relative Risk Aversion for bonds and equities in the range zero to 10. As Mehra and Prescott (1985) note, values from zero to 10 are historically reasonable in accordance with observations, e.g., Arrow (1971), Kydland and Prescott (1982), Hildreth and Knowles (1982), Evans (2005), Groom and Maddison (2019). However, the traditional view is that the utility of money is invariant and independent of context, i.e., investors do not use contextual information in estimating their aversion to risk. In contrast, our approach implies that in an efficient market, investors use all available information to maximize return on volatility, i.e. utility is polymorphic and the discount applied to each future cash flow depends on all information available on that cash flow.

# Summary

Equity premium, the surplus returns of stocks over bonds, has been an enduring puzzle. While numerous prior works approach the problem assuming the utility of money is invariant across contexts, our approach implies that in efficient markets the utility of money is polymorphic, with risk aversion dependent on the information available in each context, i.e. the discount on each future cash flow depends on all information available on that cash flow. Specifically, we prove that in efficient markets, informed investors maximize return on volatility by being risk-neutral with riskless bonds, and risk-averse with equities, thereby resolving the puzzle. We validate our results on historical data with surprising consistency.